\documentclass[twocolumn, prb]{revtex4}
\usepackage{graphicx}
\usepackage{latexsym}

\newcommand{\be}{\begin{equation}}
\newcommand{\ee}{\end{equation}}
\newcommand{\bea}{\begin{eqnarray}}
\newcommand{\eea}{\end{eqnarray}}

\begin{document}
\title{Hopping conductivity of a suspension of flexible wires in an insulator}
\author{Tao Hu, B. I. Shklovskii}
\affiliation{Department of Physics, University of Minnesota \\
116 Church Street SE, Minneapolis, MN 55455}
\date{\today}
\begin{abstract}

We study the hopping conductivity in a composite made of Gaussian
coils of flexible metallic wires randomly and isotropically
suspended in an insulator at such concentrations that the spheres
containing each wire overlap. Uncontrolled donors and acceptors in
the insulator lead to random charging of wires and, hence, to a
finite density of states (DOS) at the Fermi level. Then the Coulomb
interactions between electrons of distant wires result in the soft
Coulomb gap. At low temperatures the conductivity is due to variable
range hopping of electrons between wires and obeys the
Efros-Shklovskii (ES) law $\ln\sigma \propto -(T_{ES}/T)^{1/2}$ with
$T_{ES}$ decreasing with concentration and length of the wires. Due
to enhanced screening of Coulomb potentials, at large enough wire
concentrations and sufficiently high temperatures, the ES law is
replaced by the Mott law $\ln\sigma \propto -(T_{M}/T)^{1/4}$.

\end{abstract}
\maketitle

\section{Introduction}

The electronic transport properties of composites made of metallic
granules surrounded by some kind of insulator is widely studied.
Spheroidal metallic particles or quantum
dots~\cite{Vinokur,Gerber,Jusz}, relatively short carbon
nanotubes~\cite{Benoit,Gaal,Fuhrer,Yosida}, and long flexible
nanowires such as bundles of carbon nanotubes or conducting
polymers~\cite{Heeger1,Kaiser,Wessling} are examples of the
dispersed granules. If the concentration $n$ of the metallic
granules is large, they touch each other and the conductivity is
metallic. When $n$ is smaller than the percolation threshold $n_c$,
the composites are in the insulating side of the metal-insulator
transition and hopping is the only mechanism of conduction.

If each granule is neutral, adding or removing one electron from it
costs a considerable electrostatic energy. The charge transport is
blocked by this Coulomb charging energy and has activation
dependence of the conductivity resembling the Arrhenius law. However
it is generally observed that at low temperatures, the conductivity
$\sigma$ exhibits a temperature dependent characteristic of variable
range hopping (VRH) transport,
\be \sigma = \sigma_0\exp[-(T_0/T)^{\alpha}], \ee
where $T_0$ is a characteristic temperature. For spheroidal
particles or quantum dots~\cite{Vinokur,Gerber,Jusz}, $\alpha$ is
observed to be $1/2$. For bundles of single wall carbon
nanotubes~\cite{Benoit}, both $\alpha=1/2$ (Ref. ~\cite{Benoit}) and
$\alpha=1/4$ (Ref. ~\cite{Benoit,Gaal,Fuhrer,Yosida}) are reported.
In composites made of conducting polymers dissolved in some kind of
insulating polymer matrix~\cite{Heeger1,Kaiser,Wessling}, the index
$\alpha$ was found to be $1/2$ below the percolation threshold and
at larger concentrations, $\alpha$ evolves from $1/4$ to $1$
(Ref.~\cite{Heeger1}). The finite DOS at the Fermi level necessary
for the VRH conductivity is attributed to the charged impurities in
the insulator playing the role of randomly biased gates. Then the
interaction of electrons residing on different wires creates the
Coulomb gap, which leads to ES VRH between
granules~\cite{Zhang,Vinokur}. In our recent paper~\cite{nanotube},
we studied the puzzling crossover from ES law with $\alpha=1/2$ to
Mott law with $\alpha=1/4$ for a suspension of straight metallic
wires. We showed that the characteristic temperature $T_{ES} \propto
1/(nL^3)^2$, where $L$ is the wire length.

In this paper we would like to calculate the hopping conductivity
for the more complicated case of flexible nanowires suspended in the
insulating medium. They can be just very thin metallic wires
suspended in an insulating liquid which was abruptly frozen, or very
long single wall nanotubes or their bundles, or heavily doped
conducting polymers. Such objects can have one or many parallel
metallic channels. In the latter case, particularly for doped
conducting polymers, disorder should be important and leads to
localization of states at some distance along the
nanowire~\cite{Heeger}. Nevertheless in this paper we concentrate on
a fundamental problem of clean or weakly disordered wires.
Generalization of our theory to the case of strongly disordered
wires, where electrons are localized within distances much smaller
than the wire length, will be discussed in the conclusion.
\begin{figure}[htbp]
\includegraphics[width=0.25\textwidth]{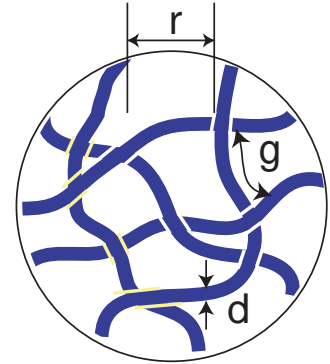}
\caption{Local view of the semidilute solutions of flexible Gaussian
wires. We show the case where the mesh size is not longer than
persistence length, the wire within each mesh is essentially
straight. At lesser density, the mesh size is longer than the
persistence length, and then the wire in the mesh is wiggly (not
shown).} \label{fig:network}
\end{figure}

We assume that each wire has the contour length $L$, relatively
large persistence length $p$ and the diameter (thickness) $d$. We
assume $L \gg p \gg d$. Throughout this work we require $L<p^3/d^2$
and this lets us disregard the effect of excluded volume,
considering each wire as a Gaussian coil with the size
$R=\sqrt{Lp}$. At small concentrations, where the inter-coil
distance is much larger than the coil size, we can simply utilize
the results for arrays of quantum dots~\cite{Zhang}. The overlap of
the coils starts when the concentration of wires exceeds the
concentration of the wire inside one coil: $n = 1/(Lp)^{3/2}$ (see
Fig. \ref{fig:network}). Therefore our discussion is confined to the
range of concentrations $1/(Lp)^{3/2} \ll n \ll n_c$, where
$n_c=1/L^2d$ is the percolation threshold at which the hoppping
conductivity is replaced by the metallic one (we derive the
expression of $n_c$ below).

In this paper we restrict ourselves to the scaling approximation for
the conductivity and to delineating the corresponding scaling
regimes. In our scaling theory, we drop away both all numerical
factors and, moreover, also all logarithmic factors, which do exist
in the problem, because it deals with strongly elongated cylinders.
In this context, we will use symbol ``$=$'' to mean ``equal up to a
numerical coefficient or a logarithmic factor''.

\begin{figure}[htb]
\includegraphics[width=0.40\textwidth]{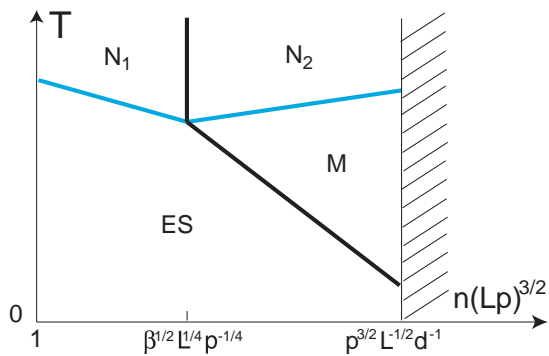}
\caption{(Color online) Summary of scaling regimes for the case of
flexible Gaussian coiled wires each with single conducting channel
and length $p^2/d<L<p^3/d^2$. Dark line separates regimes where
Coulomb interaction is important from regimes where it plays minor
role. The grey (blue) line separates activated NNH regimes $\rm N_1$
and $\rm N_2$ from VRH regimes $\rm ES$ and $\rm M$.}
\label{fig:regime-coil-1a}
\end{figure}

Our main results for the hopping conduction are presented by the
¡°phase diagrams¡± in the plane of parameters $n(Lp)^{3/2}$ and $T$
shown in Figs. \ref{fig:regime-coil-1a}, \ref{fig:regime-coil-1b}
and \ref{fig:regime-coil}. Fig. \ref{fig:regime-coil-1a} represents
the simplest case of narrow wires with only one conducting channel
and relatively long length $p^2/d<L<p^3/d^2$. Remarkably the ES
hopping regime which vanishes for single channel straight
wires~\cite{nanotube} does exist for Gaussian coiled wires. The
reason is that each straight wire is a one dimensional object where
Coulomb interaction plays only marginal role. On the other hand, a
coiled metallic wire from the point of view of the Coulomb
interaction is a two-dimensional object, where charging energy $E_c$
can be substantially larger than the quantum level spacing $\delta$.
At large enough $n(Lp)^{3/2}$ and sufficiently high temperature, ES
hopping conductivity is replaced by the Mott law due to increasing
screening of coulomb potentials (regime M). At even higher
temperatures both VRH regimes crossover to the nearest neighbor
hopping (NNH) regimes.

The plan of this article is as follows. In Sec. \ref{sec:kappa} we
first consider the macroscopic dielectric constant $\kappa_{\rm
eff}$ of the suspension. We then calculate the finite DOS at the
Fermi level in Sec. \ref{sec:DOS}. We continue in Sec. \ref{sec:xi}
by calculating the effective localization length $\xi$ for hop over
distances larger than the typical coil size. Using these results, we
summarize all the scaling regimes for the hoping conductivities in
Sec. \ref{sec:regime}. Finally in Sec. \ref{sec:discussion}, we
conclude with the discussion of generalization of our theory to the
case where wires are strongly disordered.

\section{Dielectric Constant} \label{sec:kappa}

We start with determining the static macroscopic dielectric constant
$\kappa_{\rm eff}$ of the suspension with wire concentration $1/L^3
\ll n \ll n_c$. The percolation threshold $n_c$ can be estimated
from the following argument. Percolation happens when each wire has
roughly two direct contacts with other wires. Following Ref.
\cite{Redbook}, the number of contacts per wire is of the order of
$nL^2d$ (each piece of the wire of the length $d$ has probability
$nLd^2$ to be touched). Requiring that $nL^2d$ is of the order of 1,
we obtain the percolation threshold $n_c = 1/L^2d$.~\cite{nematic}

In Ref.\cite{conductivity}, we have calculated the macroscopic
conductivity $\sigma$ of a suspension of Gaussian coiled wires with
conductivity $\sigma_1$ in a weakly conducting medium with
conductivity $\sigma_2 \ll \sigma_1$. When dealing with metal wires
in an insulator, one can consider small final frequency response and
use $-i\omega\kappa/4\pi$ for $\sigma_2$. This lets us find the
macroscopic effective conductivity $\sigma(\omega)$ and $\kappa_{\rm
eff}$ using $\kappa_{\rm eff}=4\pi i\sigma(\omega)/\omega$,
\be \kappa_{\rm eff} = \left\{ \begin{array}{lcr} \kappa n^2L^3p^3 & {\rm if} & 1 < n(Lp)^{3/2} < \frac{L^{1/2}}{p^{1/2}} \\
\\ \kappa nL^2p & {\rm if} & \frac{L^{1/2}}{p^{1/2}} < n(Lp)^{3/2} <
\frac{p^{3/2}}{L^{1/2}d}
\end{array} \label{eq:kappa} \right. \ , \ee
where $\kappa$ is the dielectric constant of the insulator. We
notice that the polarization of these metallic wires leads to great
enhancement of the macroscopic dielectric constant $\kappa_{\rm
eff}$ much larger than $\kappa$ of the insulating host. The
derivation of Eq. (\ref{eq:kappa}) is based upon the geometrical
properties of the system, therefore let us dwell on them first.

In the parlance of polymer science, when wires strongly overlap at
$n
> 1/(Lp)^{3/2}$, we deal with a semidilute solution - the system
locally looks like a network with certain mesh size $r$ (see Fig.
\ref{fig:network}). In the scaling sense, $r$ is the same as the
characteristic radius of density-density correlation function, and
can be estimated following Ref.~\cite{Redbook}. Suppose that the
wire within each mesh has a contour length $g$. It makes a density
about $g/Lr^3$ which must be about the overall average density $n$.
Thus, $g/Lr^3 = n$. Second relation between $g$ and $r$ depends on
whether mesh size is bigger or smaller than persistence length $p$:
\be r = \left\{ \begin{array}{lcr} g & {\rm if} & g < p \\
\sqrt{gp} & {\rm if} & g >p \end{array} \right. \ . \ee
Accordingly, one obtains
\be \begin{array}{lccr} g = \frac{1}{n^{2}L^2p^{3}} \ , & r =
\frac{1}{nLp} & {\rm if} & \frac{1}{(Lp)^{3/2}} < n < \frac{1}{Lp^2}
\\ g = \sqrt{\frac{1}{nL}} \ , & r = \sqrt{\frac{1}{nL}} & {\rm if}
& \frac{1}{Lp^2} < n < \frac{1}{L^2d}
\end{array}  \ , \label{eq:blob_size} \ee
The upper line corresponds to such a dense network that every mesh
is shorter than persistence length and wire is essentially straight
within each mesh (see Fig. \ref{fig:network}). The lower line
describes much less concentrated network, in which every mesh is
represented by a little Gaussian coil.

Now we present a simple derivation of Eqs. (\ref{eq:kappa})
equivalent to what we have done for the macroscopic conductivity in
Ref.\cite{conductivity}. Consider a cube of the size about
$(Lp)^{1/2}$ inside our macroscopic sample. On the one hand, the
overall dielectric constant on the scale of this cube is already
about the same as that of a macroscopic body, we denote it
$\kappa_{\rm eff}$. Therefore the capacitance of such a cube is
about $(Lp)^{1/2}\kappa_{\rm eff}$. On the other hand, we can
estimate this capacitance considering wires inside the cube. Because
the distance between wires is about $r$, each wire bridges another
wire through a blob of insulator with size $r$ and capacitance
$r\kappa$. Since the contour length of the wire inside each blob is
about $g$, there are about $L/g$ such capacitors connected in
parallel, yielding overall capacitance connecting the given wire as
$(L/g)(r\kappa)$. Since all (or sizeable fraction of all)
$n(Lp)^{3/2}$ wires in the cube are in parallel, we finally arrive
at the cube capacitance as $n(Lp)^{3/2}(L/g)(r\kappa)$. Equating
this to $(Lp)^{1/2}\kappa_{\rm eff}$ and accounting for the
different geometrical properties represented by Eqs.
(\ref{eq:blob_size}), we arrive at Eq. (\ref{eq:kappa}).

Eqs. (\ref{eq:kappa}) can also be understood along the following
way. If the wave vector $q$ is larger than $1/(Lp)^{1/2}$, the
static dielectric function for the wire suspension has the
metal-like form
\be \epsilon(q) = \kappa\left(1+\frac{1}{q^2r^{2}}\right), \ee
where $r$ represents the typical separation of the wire from other
wires and plays the role of screening radius for the wire charge.
The function $\epsilon(q)$ grows with decreasing $q$ until $q =
1/(Lp)^{1/2}$ where the composite loses its metallic response and
$\epsilon(q)$ saturates. As a result, the macroscopic effective
dielectric constant is given by $\kappa_{\rm eff} = \epsilon(q =
1/(Lp)^{1/2})$. Plugging proper $r$ at different concentrations (Eq.
(\ref{eq:blob_size})), we obtain $\kappa_{\rm eff}$ represented by
Eqs. (\ref{eq:kappa}). They show how $\kappa_{\rm eff}$ grows with
$L$ at $L \gg p$ for flexible nanowires. In the following sections,
we will use Eqs. (\ref{eq:kappa}) for the calculation of the hopping
conductivity. Note, however, that the dielectric constant of our
system may be itself of interest in applications where very large
dielectric constants are welcomed.

\section{Density of States} \label{sec:DOS}

As we mentioned in the introduction, according to Ref.~\cite{Zhang},
the finite density of states (DOS) near the Fermi level originates
from uncontrollable donors (or acceptors) in the insulating host.
Donors have the electron energy above the Fermi energy of wires.
Therefore, they donate electrons to wires. A positively charged
ionized donor can attract and effectively bind fractional negative
charges on all neighboring wires, leaving the rest of each wire
fractionally charged. At a large enough average number ($\gg 1$) of
donors per wire, effective fractional charges on different wires are
uniformly distributed from $-e/2$ to $e/2$. In such a way the
Coulomb blockade in a single wire is lifted and the discrete density
of states get smeared. If the charging energy $E_c$ for a coil is
larger than the mean quantum level spacing $\delta$ inside the wire,
the DOS $g_0$ is given by $g_0 = 1/E_c(Lp)^{3/2}$. Substituting the
effective dielectric constants given by Eqs. (\ref{eq:kappa}), we
obtain the charging energy
\be E_c = \left\{ \begin{array}{lcr} e^2/(\kappa n^2L^{7/2}p^{7/2}) & {\rm if} & 1 < n(Lp)^{3/2} < \frac{L^{1/2}}{p^{1/2}} \\
\\ e^2/(\kappa nL^{5/2}p^{3/2}) & {\rm if} & \frac{L^{1/2}}{p^{1/2}} < n(Lp)^{3/2} <
\frac{p^{3/2}}{L^{1/2}d}
\end{array} \label{eq:EC} \right. \ . \ee
In the very vicinity of the Fermi energy, the long range Coulomb
interaction creates the parabolic Coulomb gap $\Delta$ which leads
to the ES VRH.

If $E_c$ is smaller than the mean quantum level spacing $\delta$,
the Coulomb interaction plays minor role. The density of states
$g_0$ becomes $1/\delta(Lp)^{3/2}$. Level spacing $\delta$ strongly
depends on the cross section of the wire. If the wire is so narrow
that it has only one conducting channel, $\delta$ is the order of
$\hbar^2/m\lambda_FL$, where $\lambda_F$ is the Fermi wavelength and
$m$ is the effective electron mass. In order to compare with the
charging energy $E_c$, we introduce constant $\beta =\lambda_F/a$
and rewrite $\delta$ as $e^2/\beta\kappa L$. Most likely, $\beta$ is
the order of $1$. If the wire is thick enough, the electrons have
three dimensional energy dispersion and $\delta$ is the order of
$\lambda_F\hbar^2/mLd^2$ or $\beta e^2 a^2/\kappa d^2L$. For both
cases, the ratio $E_c/\delta$ decreases with growing $n$ or $L$
because of enhanced screening of Coulomb interaction in a denser
system of longer wires. To calculate the VRH conductivity, we need
now another key quantity--the localization length $\xi$.

\section{Localization Length} \label{sec:xi}

It was suggested~\cite{Boris} that long range hopping process
exceeding the size of a single wire can be realized by tunneling
through a sequence of wires. The states of the intermediate wires
participate in the tunneling process as \emph{virtual states}. The
virtual electron tunneling through a single granule is called
co-tunneling~\cite{Nazarov,Ioselevich,Beloborodov,Vinokur,Fogler1}
and regarded as a key mechanism of low temperature charge transport
via quantum dots. One should distinguish the two co-tunneling
mechanisms, elastic and inelastic. During the process of elastic
co-tunneling, the electron tunneling through an intermediate virtual
state in the granule leaves the granule with the same energy as its
initial state. On the contrary, the tunneling electron in the
inelastic co-tunneling mechanism leaves the granule with an excited
electron-hole pair behind it. Which mechanism dominates the
transport depends on the temperature. Inelastic co-tunneling
dominates at $T>\sqrt{E_c\delta}$, while below this temperature,
elastic co-tunneling wins. In discussions below, we consider
sufficiently low temperatures such that only elastic co-tunneling is
involved.

Let us remind the idea of the calculation of the localization length
$\xi$ used in Ref.~\cite{Boris,Zhang,Fogler}. When an electron
tunnels through the insulator between wires at the nearest
neighboring hopping (NNH) distance $r_n$, it accumulates
dimensionless action $r_n/a$, where $a$ is the tunneling decay
length in the insulator. We want to emphasize that the NNH distance
$r_n$ is realized only in one point of the wire and, therefore, is
much shorter than the typical separation along the wire from other
wires $r$. NNH distance $r_n$ can be calculated using the
percolation method~\cite{ES}. If one consider each wire with
thickness $r_n$, using the same idea we showed above, percolation
through these cylinders appears, when $nL^2r_n = 1$. Thus $r_n =
1/nL^2$. Since we assume the metallic wire is only weakly
disordered, we neglect the decay of electron wave function in the
wire. Over distance $x$, electron tunnels $x/(Lp)^{1/2}$ times
accumulating a dimensionless action of the order of $r_n/a$ each
time. Thus, its wave function decays as
$\exp[-xr_n/(Lp)^{1/2}a]=\exp(-x/\xi)$, where the localization
length
\be \xi=a(Lp)^{1/2}/r_n=a(nL^{5/2}p^{1/2}). \label{eq:xi} \ee
Now we have all the elements needed to construct our phase diagrams
of the conductivity at various temperatures and concentrations.

\section{Summary of Phase Diagrams} \label{sec:regime}

\begin{figure}
\includegraphics[width=0.40\textwidth]{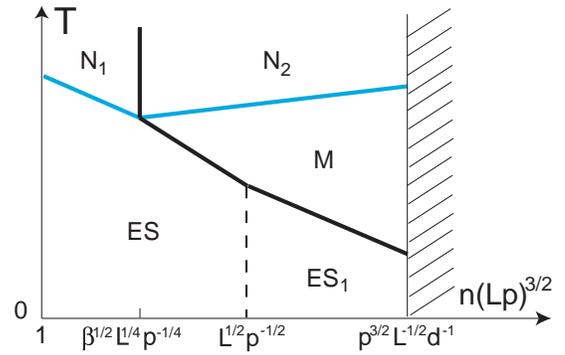}
\caption{(Color online) Summary of scaling regimes for the case of
flexible Gaussian coiled wires each with single conducting channel
and length $p<L<p^2/d$. The dashed line marks the border where the
wire within each mesh changes from Gaussian coiled to be straight.
The effective dielectric constant $\kappa_{\rm eff}$ also changes at
this line. } \label{fig:regime-coil-1b}
\end{figure}

We first consider narrow wires each with only one conducting
channel. Depending on the relation between $L$ and $p$, percolation
can start before or after $n(Lp)^{3/2} = (L/p)^{1/2}$, where the
Gaussian coiled wire within each mesh becomes straight at higher
concentrations. Our results for these two cases are summarized in
Figs. \ref{fig:regime-coil-1a} and \ref{fig:regime-coil-1b} in the
plane of parameters $n(Lp)^{3/2}$ and $T$. Each line represents a
smooth crossover except the percolation threshold $n(Lp)^{3/2} =
p^{3/2}/L^{1/2}d$, where the insulator-metal transition happens. In
the vicinity of this transition, the conductivity has a critical
behavior, which goes beyond the scope of our scaling approach.

To be systematic, let us start the review of scaling regimes from
Fig. \ref{fig:regime-coil-1a}, which is valid when $p^2/d<L<p^3/d^2$
and therefore percolation happens at relatively dilute solution with
Gaussian coiled wire inside each mesh. We begin with activated NNH
hopping regimes $\rm N_1$ and $\rm N_2$ at sufficiently high
temperatures. The temperature dependence of the conductivity is
given by
\be \sigma \propto
\exp\left(-\frac{1}{nL^2a}\right)\exp\left(-\frac{E_A}{T}\right),
\label{eq:NNH} \ee
where the first term is due to electron wave function overlap and in
the second term $E_A = max\{E_c, \delta\}$ is the activation energy
at very small tunneling conductances. Here the charging energy $E_c$
should be taken from the upper line of Eq. (\ref{eq:EC}). Due to
enhanced screening of Coulomb potentials with growing $n(Lp)^{3/2}$,
$E_c$ decreases. As a result, the activation energy $E_A$ evolves
from $E_c$ in regime $\rm N_1$ to the quantum level spacing $\delta$
in regime $\rm N_2$ at $n(Lp)^{3/2}=\beta^{1/2}L^{1/4}/p^{1/4}$. The
Eq. (\ref{eq:NNH}) is valid as far as the contribution of the
activation is smaller than that of the overlap. When they are
comparable at sufficiently low temperatures, the NNH regimes
smoothly crossover to VRH regimes at $T=e^2a/\kappa nL^{3/2}p^{7/2}$
and $T=e^2anL/\beta\kappa$ for the regimes $\rm N_1$ and $\rm N_2$
respectively.
\begin{figure}
\includegraphics[width=0.40\textwidth]{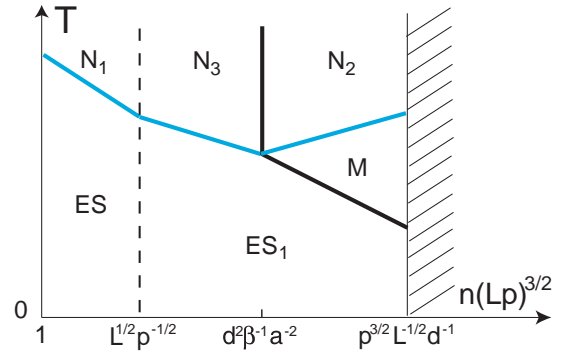}
\caption{(Color online) Summary of scaling regimes for the case of
thick wires with length $p<L<p^2/d$.} \label{fig:regime-coil}
\end{figure}

In the regime $\rm ES$, Coulomb interaction is important, which
leads to the Efros-Shklovskii (ES) VRH conductivity
\be \sigma = \sigma_0\exp[-(T_{ES}/T)^{1/2}], \label{eq:ES}\ee
where $T_{ES}=e^2/\kappa_{\rm eff}\xi$. Applying the above line of
Eqs. (\ref{eq:kappa}) and Eq. (\ref{eq:xi}), we obtain
\be T_{ES} = \frac{e^2}{\kappa n^3L^{11/2}p^{7/2}a}. \ee

At large $n(Lp)^{3/2}$ and relatively high temperatures, the ES law
is replaced by the conventional Mott law,
\be \sigma = \sigma_0\exp[-(T_M/T)^{1/4}], \label{eq:Mott}\ee
where $T_M = 1/g_0\xi^3$. Plugging in $g_0$ for the narrow wires and
Eq. (\ref{eq:xi}), we arrive at
\be T_M=\frac{e^2}{\beta \kappa n^3L^7a^3}.\ee
The regime $\rm ES$ borders the regime M along the line $T=\beta
e^2a/\kappa n^3L^4p^7$, which can be obtained by equating the
Coulomb gap width $\Delta=\sqrt{g_0e^6/\kappa_{\rm eff}^3}$ to the
Mott's optimal band width $T^{3/4}/(g_0\xi^3)^{1/4}$.

When the wire is so short that $p<L<p^2/d$, percolation happens at
relatively dense system where the wire within each mesh is straight.
As a result, to the right of the transition from the coiled mesh to
the straight mesh at $n(Lp)^{3/2}=(L/p)^{1/2}$, the effective
dielectric constant should be taken from the bottom line of Eq.
(\ref{eq:kappa}). Thus for the regime $\rm ES_1$, the characteristic
temperature $T_{ES}$ in Eq. (\ref{eq:ES}) should be replaced by
\be T_{ES_1} = \frac{e^2}{\kappa an^2L^{9/2}p^{3/2}}. \ee
The crossover from this new regime $\rm ES_1$ to the regime M is
along the line $T=\beta e^2a/\kappa nL^2p^3$. It is worthwhile to
emphasize that Coulomb interaction loses its importance for the NNH
regimes at $n(Lp)^{3/2}=\beta^{1/2}(L/p)^{1/4}$ which is smaller
than $(L/p)^{1/2}$, where the geometry of the mesh changes and thus
so does the effective dielectric constant $k_{\rm eff}$. As a
result, compared to Fig. \ref{fig:regime-coil-1a}, the regime $\rm
ES_1$ is the only new regime added in Fig. \ref{fig:regime-coil-1b}.
At $p=L$, Fig. \ref{fig:regime-coil-1b} resembles the case of
straight wires, the ES regimes almost vanish. Until now, we complete
the review of phase diagrams for the narrow wires each with single
conducting channel.

For thick wires, simply using the corresponding density of states
discussed in the Sec. \ref{sec:DOS}, we obtain phase diagrams
similar to those for the case of narrow wires, but with new borders
between scaling regimes. Unlike narrow wires, there is a possibility
that the geometry of the mesh changes when the Coulomb interaction
is still important. Scaling regimes of this case is shown in Fig.
\ref{fig:regime-coil}. There is a new NNH regime $\rm N_3$ where the
the conductivity is represented by Eq. (\ref{eq:NNH}) with
activation energy $E_A$ given by the bottom line of Eq.
(\ref{eq:EC}). Replacing $p$ by $L$ in Fig. \ref{fig:regime-coil},
we again recover the result we got for straight wires in Fig. 2 of
Ref. \cite{nanotube}.

\section{Conclusion} \label{sec:discussion}

In this paper we calculated the macroscopic dielectric constant and
the hopping conductivity of a suspension of Gaussian coiled metallic
wires in an insulator. We obtained a plethora of different scaling
regimes. Our results are applicable to wires made of well conducting
metals with relatively small disorder. They can also be generalized
to the case of strongly disordered wires where electrons are
localized at a distance $\lambda$ much smaller than the length of
the wire $L$. To this end, one should proceed similar to what we
have done for straight wires in the Ref.\cite{nanotube}. To
calculate $\kappa_{\rm eff}$, one can imagine that the disorder
effectively \emph{cuts} the wire into shorter metallic pieces each
with length $\lambda$. Therefore, we can calculate $\kappa_{\rm
eff}$ thinking about a suspension of metal wires with shorter
lengthes $\lambda$ but larger concentration $nL/\lambda$. Resulting
$\kappa_{\rm eff}$ depends on $\lambda$ and is much smaller than
that of clean wires. Calculating the effective localization length
$\xi$ similarly to Sec. \ref{sec:xi}, we should add the decay of
electron wave functions along the wire. Taking into account the
fractal properties of the wire, we arrive at a much smaller
localization length than that of clean metal. Reduction of
$\kappa_{\rm eff}$ and $\xi$ leads to much larger $T_{ES}$, $T_M$
and much smaller hopping conductivity. We can construct phase
diagrams which topologically look similar to diagrams of Figs.
\ref{fig:regime-coil-1a}, \ref{fig:regime-coil-1b} and
\ref{fig:regime-coil}, but they are more complicated and we are not
showing them here.

\acknowledgements We are grateful to M. M. Fogler, A. Yu. Grosberg
for useful discussions.


\begin{thebibliography}{99}
\bibitem{nematic} The nematic ordering of wires starts at a larger concentration $n
= 1/Lpd$. Percolation starts earlier than the nematic ordering,
because percolation requires a couple of contacts per wire, while
nematic ordering requires a contact per a smaller persistance length
$p$.
\end{thebibliography}
\end{document}